\documentclass[12pt]{article}

\def\ov{\overline}

\def\la{\langle}
\def\ra{\rangle}
\def\be{\begin{equation}}
\def\ee{\end{equation}}

\newtheorem{theorem}{Theorem}
\newtheorem{lemma}[theorem]{Lemma}

\newtheorem{definition}[theorem]{Definition}

\begin{document}

\author{S.Albeverio, S.V.Kozyrev}

\title{Coincidence of the continuous and discrete \\ $p$--adic wavelet transforms}

\maketitle

\begin{abstract}
We show that translations and dilations of a $p$--adic wavelet
coincides (up to the multiplication by some root of one) with a
vector from the known basis of discrete $p$--adic wavelets. In this
sense the continuous $p$--adic wavelet transform coincides with the
discrete $p$--adic wavelet transform.

The $p$--adic multiresolution approximation is introduced and
relation with the real multiresolution approximation is described.

Relation of application of $p$--adic wavelets to spectral theory of
$p$--adic pseudodifferential operators and the known results about
sparsity of matrices of some real integral operators in the bases of
multiresolution wavelets is discussed.

\end{abstract}


multiresolution analysis

\bigskip


\section{Introduction}

Wavelet analysis (see e.g. \cite{Daubechies}, \cite{Meyer}) is an
important method used in a wide variety of applications from the
theory of functions to signal analysis. For the standard wavelet
analysis of functions of a real argument one can consider both the
continuous wavelet transform and the discrete wavelet transform
(expansion over wavelet bases).

A basis of wavelets in spaces of complex valued functions of a
$p$--adic argument was introduced and its relation to the spectral
analysis of $p$--adic pseudodifferential operators was investigated
in \cite{wavelets}. One can also consider, in analogy with the real
case, a continuous $p$--adic wavelet transform, see \cite{Altaisky}
for a discussion.

In the present paper we show that in the $p$--adic case, with the
proper choice of a wavelet, an arbitrary translation and dilation
gives (up to the multiplication by some root of one) some vector
from the basis of discrete $p$--adic wavelets introduced in
\cite{wavelets}. In this sense the continuous $p$--adic wavelet
transform coincides with the discrete $p$--adic wavelet transform.

This result looks strange from the point of view of real analysis.
The observed behavior is a manifestation of the ultrametricity of
the $p$--adic norm. In an ultrametric space it is not possible to
leave a ball making steps smaller than the diameter of this ball.
This implies the existence of locally constant functions
--- functions which are constant on some vicinity for any point but
not necessarily constant globally.

When we apply the transformation of dilation or translation to a
locally constant function of a $p$--adic argument, then, if the
transformation is sufficiently small, the function will be
invariant. This allows us to expect some properties of discreteness
for continuous $p$--adic wavelet transform. In the present paper we
show that making the right choice of wavelet it is possible to make
continuous and discrete $p$--adic wavelet transforms to be equal.

The next construction of the present paper is the multiresolution
approximation of $L^2(Q_p)$. Unlike in the real case, where
translations by elements of the subgroup of integers in the real
line were used, in the definition of the $p$--adic multiresolution
approximation we use translations by elements of the factorgroup
$Q_p/Z_p$. We show that the Bruhat--Schwartz space ${\cal D}(Q_p)$
of $p$--adic test functions can be considered as a multiresolution
approximation of $L^2(Q_p)$. In this sense the construction of
multiresolution analysis (MRA) is a natural property of the
$p$--adic analysis.

These results, together with the simple relation of $p$--adic
wavelets to spectral theory of $p$--adic pseudodifferential
operators described in \cite{wavelets} (the real analogue of this
relation we will discuss in Section 3), supports the point of view
that the wavelet analysis is, in essence, a $p$--adic theory. In the
real case we obtain a much more complicated wavelet analysis
compared to the $p$--adic case.

The structure of the present paper is as follows.

In Section 2 we consider the continuous $p$--adic wavelet transform
and show that it coincides with the discrete wavelet transform.

In Section 3 we introduce the $p$--adic multiresolution
approximation and discuss its relation with the real multiresolution
approximation. We also discuss relation of application of $p$--adic
wavelets to spectral theory of $p$--adic pseudodifferential
operators and the known results about sparsity of matrices of some
real integral operators in the bases of multiresolution wavelets.

In Section 4, the Appendix, we put the material on $p$--adic
analysis and $p$--adic wavelets, used in the present paper.

\section{Continuous $p$--adic wavelet transform}

For the notations used in the present section see the Appendix.

Let us consider the main construction of the present paper --- the
continuous $p$--adic wavelet transform generated by the $p$--adic
wavelet of the form
$$
\psi(x)=\chi(p^{-1}x)\Omega(|x|_p)
$$
which is equal to the product of a character $\chi$ and the
characteristic function $\Omega$ of a ball.

In the next lemma we show that any vector from the basis
$\{\psi_{\gamma nj}\}$ of $p$--adic wavelets is a translation and
dilation of the wavelet $\psi$, and, vice versa, any translation and
dilation of the wavelet $\psi$ belongs to the basis $\{\psi_{\gamma
nj}\}$ up to multiplication by some root of one.

Equivalently, the basis of $p$--adic wavelets (multiplied by
corresponding roots of one) can be considered as an orbit of action
of the $p$--adic affine group in $L^2(Q_p)$ containing the function
$\psi$.

\begin{lemma}\label{Lem2}\qquad {\sl 1) Any wavelet from the basis $\{\psi_{\gamma
nj}\}$ of $p$--adic wavelets is a translation and dilation of the
wavelet
$$
\psi(x)=\chi(p^{-1}x)\Omega(|x|_p)
$$
of the following form
$$
\psi_{\gamma nj}(x)=\psi^{p^{-\gamma}j^{-1},p^{-\gamma}n}(x)
$$

\noindent 2) The translation and dilation of the $p$--adic wavelet
$\psi$ is proportional to a vector from the basis $\{\psi_{\gamma
nj}\}$: \be\label{tra_dil}
\psi^{a,b}(x)=\chi\left(p^{-1}\left[(a|a|_p)^{-1}\,{\rm
mod}\,p\right](\{|a|_pb\}-[|a|_pb\,{\rm mod}\,p])\right)\psi_{\log_p
|a|_p,\, \{|a|_pb\}   ,\,(a|a|_p)^{-1}\,{\rm mod}\,p} \ee Here
$a,b\in Q_p$, $a\ne 0$.

}

\end{lemma}

\noindent{\it Proof}\qquad 1) One has
$$
\psi_{\gamma nj}(x)=p^{-{\gamma\over 2}} \chi(p^{-1}j
(p^{\gamma}x-n)) \Omega(|p^{\gamma} x-n|_p)=
$$
$$
=p^{-{\gamma\over 2}} \chi(p^{-1}(p^{\gamma}jx-nj))
\Omega(|p^{\gamma} jx-nj|_p)=p^{-{\gamma\over
2}}\psi(p^{\gamma}jx-nj)=\psi^{p^{-\gamma}j^{-1},p^{-\gamma}n}(x)
$$

2) Let us compute
$$
\psi^{a,b}(x)=|a|_p^{-{1\over 2}}\psi\left({x-b\over
a}\right)=|a|_p^{-{1\over 2}}\chi\left(p^{-1}{x-b\over
a}\right)\Omega\left(\left|{x-b\over a}\right|_p\right)
$$
We have
$$
a^{-1}=|a|_p {a^{-1}\over |a|_p},\qquad \Omega\left(\left|{x-b\over
a}\right|_p\right)=\Omega\left(||a|_px-|a|_pb|_p\right)=\Omega\left(||a|_px-\{|a|_pb\}|_p\right)
$$
$$
\chi\left(p^{-1}{x-b\over a}\right)\Omega\left(\left|{x-b\over
a}\right|_p\right)=\chi\left(p^{-1}{a^{-1}\over
|a|_p}(|a|_px-|a|_pb)\right)\Omega\left(||a|_px-\{|a|_pb\}|_p\right)=
$$
$$
=\chi\left(p^{-1}\left[(a|a|_p)^{-1}\,{\rm
mod}\,p\right](|a|_px-|a|_pb)\right)\Omega\left(||a|_px-\{|a|_pb\}|_p\right)=
$$
$$
=\chi\left(p^{-1}\left[(a|a|_p)^{-1}\,{\rm
mod}\,p\right](\{|a|_pb\}-[|a|_pb\,{\rm mod}\,p])\right)
$$
$$
\chi\left(p^{-1}\left[(a|a|_p)^{-1}\,{\rm
mod}\,p\right](|a|_px-\{|a|_pb\})\right)\Omega\left(||a|_px-\{|a|_pb\}|_p\right)
$$
Taking into account the normalization, we get
$$
\psi^{a,b}(x)=\chi\left(p^{-1}\left[(a|a|_p)^{-1}\,{\rm
mod}\,p\right](\{|a|_pb\}-[|a|_pb\,{\rm mod}\,p])\right)\psi_{\log_p
|a|_p,\, \{|a|_pb\} ,\,(a|a|_p)^{-1}\,{\rm mod}\,p}
$$
i.e. the above translation and dilation of the wavelet $\psi$
coincides with a wavelet from the basis of $p$--adic wavelets times
a root of one. This finishes the proof of the lemma.

\bigskip

The next theorem shows that, in the $p$--adic case, the continuous
wavelet theory in essence coincides with the discrete wavelet
theory.

\begin{theorem}\qquad{\sl
1) The $p$--adic wavelet transform $T$, corresponding to the wavelet
$\psi$, takes the form of an expansion over the basis of $p$--adic
wavelets $\psi_{\gamma nj}$:
$$
(Tf)(a,b)=|a|_p^{-{1\over 2}} \int dx f(x)\ov{\psi}\left({x-b\over
a}\right)=\la \psi^{a,b},f\ra=
$$
$$
=\chi\left(-p^{-1}\left[(a|a|_p)^{-1}\,{\rm
mod}\,p\right](\{|a|_pb\}-[|a|_pb\,{\rm mod}\,p])\right)f_{\log_p
|a|_p,\, \{|a|_pb\} ,\,(a|a|_p)^{-1}\,{\rm mod}\,p}
$$
where  $f_{\gamma n j}$ is the coefficient of expansion of the
function $f$ over the basis of wavelets.

\noindent 2) For the continuous $p$--adic wavelet transform related
to the $p$--adic wavelet $\psi$ the formula for the inverse wavelet
transform takes the form of the expansion of the unit operator over
the basis of $p$--adic wavelets
$$
C_{\psi}^{-1}\int{dadb\over |a|_p^2} |\psi^{a,b}\ra
\la\psi^{a,b}|=\sum_{\gamma\in Z,n\in Q_p/Z_p,j=1,\dots,p-1}
|\psi_{\gamma n j}\ra \la \psi_{\gamma n j}|=1
$$

}

\end{theorem}

\noindent{\it Proof}\qquad  The first statement follows directly
from Lemma \ref{Lem2}.

Let us compute for the $p$--adic wavelet $\psi$ the constant
$$
C_{\psi}={1\over\|\psi\|^2}\int{dadb\over
|a|_p^2}|\la\psi(x),\psi^{ab}(x)\ra|^2=p^{-1}
$$
Here we used (\ref{tra_dil}) and the orthogonality of the wavelets
in the basis, $\|\psi\|$ is the $L^2$--norm.

Then, using (\ref{tra_dil}), we compute the integral
$$
\int{dadb\over |a|_p^2} |\psi^{a,b}\ra \la\psi^{a,b}|=\int{dadb\over
|a|_p^2} |\psi_{\log_p |a|_p,\, \{|a|_pb\} ,\,(a|a|_p)^{-1}\,{\rm
mod}\,p}\ra\la \psi_{\log_p |a|_p,\, \{|a|_pb\}
,\,(a|a|_p)^{-1}\,{\rm mod}\,p}|=
$$
$$
=\sum_{n\in Q_p/Z_p}\int{da\over |a|_p} |\psi_{\log_p |a|_p,\, n
,\,(a|a|_p)^{-1}\,{\rm mod}\,p}\ra\la \psi_{\log_p |a|_p,\, n
,\,(a|a|_p)^{-1}\,{\rm mod}\,p}|=p^{-1}\sum_{\gamma\in Z,n\in
Q_p/Z_p,j=1,\dots,p-1} |\psi_{\gamma n j}\ra \la \psi_{\gamma n j}|
$$
Therefore
$$
C_{\psi}^{-1}\int{dadb\over |a|_p^2} |\psi^{a,b}\ra
\la\psi^{a,b}|=\sum_{\gamma\in Z,n\in Q_p/Z_p,j=1,\dots,p-1}
|\psi_{\gamma n j}\ra \la \psi_{\gamma n j}|=1
$$
This finishes the proof of the theorem.

\bigskip

\section{Multiresolution approximation of $L^2(Q_p)$ and integral operators}

The following definition of a multiresolution approximation can be
found in \cite{Meyer}.

\begin{definition}\qquad {\sl A multiresolution approximation of
$L^2(R)$ is a decreasing sequence $V_{\gamma}$, $\gamma\in Z$, of
closed linear subspaces of $L^2(R)$ with the following properties:
1)
$$
\bigcap_{-\infty}^{+\infty}V_{\gamma}=\{0\},\qquad
\bigcup_{-\infty}^{+\infty}V_{\gamma}\quad {\rm is~dense~in}\quad
L^2(R);
$$
2)  for all $f\in L^2(R)$ and all $\gamma\in Z$,
$$
f(\cdot)\in V_{\gamma}  \Longleftrightarrow  f(2\cdot)\in V_{j-1};
$$
3) for all $f\in L^2(R)$ and all $n\in Z$,
$$
f(\cdot)\in V_0 \Longleftrightarrow f(\cdot-n)\in V_0;
$$
4)  there exists a function $g\in V_0$ such that the sequence
$$
g(\cdot-n),\quad n\in Z
$$
is a Riesz basis of the space $V_0$.

}

\end{definition}

It follows that there exists a function $\phi\in V_0$, such that the
sequence $\phi(x-n)$, $n\in Z$ is an orthonormal basis of the space
$V_0$. In the following we will speak only about orthonormal bases.

We denote by $W_\gamma$ the orthogonal complement to $V_\gamma$ in
the space $V_{\gamma-1}$. Then $L^2(R)$ is the orthogonal sum of the
$W_\gamma$:
$$
L^2(R)=\oplus_{\gamma\in Z} W_{\gamma}
$$
There exists a well known construction which allows to build wavelet
bases from multiresolution approximations, see \cite{Daubechies} or
\cite{Meyer} for the details. The space $W_\gamma$ possesses the
basis $\{\Psi_{\gamma n}\}$, $n\in Z$ of multiresolution wavelets.

\bigskip

\noindent{\bf Example 1}\qquad To define a multiresolution
approximation it is sufficient to fix the function $\phi$. Let us
take this function to be equal to the characteristic function of the
interval $[0,1]$:
$$
\phi(x)=\chi_{[0,1]}(x),\qquad x\in R
$$
This choice of function $\phi$ is related to the Haar wavelets. It
is easy to see that for the example under consideration all the
properties of multiresolution approximation will be satisfied.
Taking $\phi(x)=\chi_{[0,1]}(x)$ and $n\ge 0$ in the definition
above, we obtain the multiresolution approximation for the space
$L^2(R_{+})$ of quadratically integrable functions on the positive
half--line.

\bigskip

It is easy to find a $p$--adic analogue of the multiresolution
approximation. The main difference is that in the $p$--adic case we
should use translations by elements of the factorgroup $Q_p/Z_p$
instead of translations by integers (we represent $n\in Q_p/Z_p$ by
the numbers (\ref{nQpZp}), see the Appendix).

\begin{definition}\label{padicMRA}\qquad {\sl A multiresolution approximation of
$L^2(Q_p)$ is a decreasing sequence $V_\gamma$, $\gamma\in Z$, of
closed linear subspaces of $L^2(Q_p)$ with the following properties:
1)
$$
\bigcap_{-\infty}^{+\infty}V_{\gamma}=\{0\},\qquad
\bigcup_{-\infty}^{+\infty}V_{\gamma}\quad {\rm is~dense~in}\quad
L^2(Q_p);
$$
2)  for all $f\in L^2(Q_p)$ and all $\gamma\in Z$,
$$
f(\cdot)\in V_{\gamma}  \Longleftrightarrow  f(p^{-1}\cdot)\in
V_{\gamma-1};
$$
3) for all $f\in L^2(Q_p)$ and all $n\in Q_p/Z_p$,
$$
f(\cdot)\in V_0 \Longleftrightarrow f(\cdot-n)\in V_0;
$$
4)  there exists a function $\phi\in V_0$ such that the sequence
$$
\phi(\cdot-n),\quad n\in Q_p/Z_p
$$
is an orthonormal basis of the space $V_0$.

}

\end{definition}

We do not speak here about the Riesz bases since, as we will see, in
the $p$--adic case we have a natural example of orthonormal basis
satisfying the above properties.

The space ${\cal D}(Q_p)$ of $p$--adic test functions possesses a
natural filtration. Let us denote by ${\cal D}_{\gamma}(Q_p)$ the
space of locally constant compactly supported functions with the
diameter of local constancy $p^{\gamma}$:
$$
|x-y|_p\le p^{\gamma}\Rightarrow f(x)=f(y)
$$
We have the following theorem, which shows that the space $L^2(Q_p)$
possesses the natural multiresolution approximation by the space
${\cal D}(Q_p)$ of test functions. Therefore the multiresolution
approximation can be considered as an intrinsic property of spaces
of functions of a $p$--adic argument. Moreover, the $p$--adic change
of variable (\ref{padicchange}) maps the multiresolution
approximation of $L^2(Q_p)$ onto the multiresolution approximation
of $L^2(R_+)$.

\begin{theorem}\qquad {\sl 1) One has the filtration
$$
{\cal D}(Q_p)= \bigcup_{-\infty}^{+\infty}{\cal D}_{\gamma}(Q_p);
$$

\noindent 2) The sequence ${\cal D}_{\gamma}(Q_p)$, $\gamma\in Z$ is
a multiresolution approximation of $L^2(Q_p)$. For this
multiresolution approximation the function $\phi(x)$ can be taken to
be equal to the characteristic function of the unit ball:
$$
\phi(x)=\Omega(|x|_p)
$$

\noindent 3) For $p=2$ the map (\ref{padicchange}) maps the
multiresolution approximation $\{V_{\gamma}\}$ of $L^2(R_+)$
described in the Example 1 onto the multiresolution approximation
$\{{\cal D}_{\gamma}(Q_p)\}$ of $L^2(Q_p)$.

The corresponding isomorphism of spaces ${\cal D}_{\gamma}(Q_p)$ and
$V_{\gamma}$ is given by the following formula:
$$
\Omega(|p^{\gamma}x-n|_p)=\chi_{[0,p^{\gamma}]}(\rho(x)-p^{\gamma}\rho(n)),\qquad
x\in Q_p,\quad n\in Q_p/Z_p.
$$

}

\end{theorem}

\noindent {\it Proof}\qquad The first statement of the theorem means
that for any locally constant function with compact support there
exists a nonzero infimum of diameters of local constancy. This
follows from the compactness of the support.

The proof of the second statements of the theorem is
straightforward. The function $\phi$ in the Definition
\ref{padicMRA} can be taken equal to be to the characteristic
function of the ball:
$$
\phi(x)=\Omega(|x|_p)
$$
The translations $\Omega(|x-n|_p)$, $n\in Q_p/Z_p$ given by
(\ref{nQpZp}), provide an orthonormal basis in ${\cal D}_0(Q_p)$.
Dilations by the degrees of $p$ give the spaces ${\cal
D}_{\gamma}(Q_p)$.

The third statement follows from Lemma \ref{ballmaps} and the above
choice of the function $\phi$. The space ${\cal D}_{\gamma}(Q_p)$
possesses the basis $\{\Omega(|p^{\gamma}x-n|_p)\}$, $x\in Q_p$,
$n\in Q_p/Z_p$. The set of functions
$\{\chi_{[0,p^{\gamma}]}(x-p^{\gamma}m)\}$, $m\in N_0$, $x\in R_{+}$
constitutes a basis in $V_{\gamma}$. Since the map $\rho$ is a
one-to-one correspondence between $Q_p/Z_p$ and $N_0$, this finishes
the proof of the theorem.

\bigskip

Note that, contrary to the real case, where characteristic functions
and Haar wavelets are discontinuous, the multiresolution
approximation of $L^2(Q_p)$ considered in the above theorem consists
of spaces of continuous functions. After the $p$--adic change of
variable the construction which was not regular (was discontinuous)
in the real case, becomes highly regular in the $p$--adic case.

\bigskip

Discuss now application of wavelets to investigation of integral
operators. It was found \cite{BCR} that for some families of
integral operators, say the Calderon--Zygmund and the
pseudodifferential operators, matrices of these operators will be
close to diagonal in wavelet bases for which wavelets has many
momenta equal to zero. Consider the multiresolution approximation
for $L^2(R)$ for which the corresponding wavelet has $M$ vanishing
moments:
$$
\int \psi(x)x^mdx=0,\qquad m=0,1,\dots M-1.
$$

Then \cite{BCR} we have the following property of decay for matrix
elements of some integral operators
$$
|\alpha^{\gamma}_{il}|+|\beta^{\gamma}_{il}|+|\gamma^{\gamma}_{il}|\le
{C_M\over 1+|i-l|^{M+1}}
$$
for all
$$
|i-l|\ge 2M
$$
Here the matrix elements $\alpha^{\gamma}_{il}$,
$\beta^{\gamma}_{il}$, $\gamma^{\gamma}_{il}$ are taken in the
wavelet bases in the spaces $V_{\gamma}$ and $W_{\gamma}$ of the
multiresolution approximation, the index $\gamma$ is related to the
scale of the wavelets and the indices $i$, $l$ related to space
localization (i.e. these are translation indices) of the wavelets.

This means that matrices of the corresponding integral operators in
the wavelet bases are sparse and close to diagonal matrices. This
result has a variety of applications, for example to numerical
algorithms of fast multiplication of integral operators in wavelet
bases.

Discuss now the $p$--adic analogue of this result. It is known
\cite{wavelets}, see the Theorem \ref{wav1} in the Appendix, that
$p$--adic pseudodifferential operators are diagonal in the basis of
$p$--adic wavelets. Moreover, in the $p$--adic case we do not need
any conditions of vanishing of higher moments for the wavelets.
Therefore in the $p$--adic case we have much stronger results about
the relation of wavelet bases and integral operators. The
diagonality of matrices of integral operators which was approximate
for the real case becomes exact for the $p$--adic case, and,
moreover, we are able to compute the spectra of the corresponding
operators.

\section{Appendix}

Let us recall some constructions of $p$--adic analysis, see
\cite{VVZ} for details and, e.g. \cite{RandomWalk}, \cite{Andr} for
applications. The field $Q_p$ of $p$--adic numbers is the completion
of $Q$ with respect to the $p$--adic norm $|\cdot|_p$, defined as
follows. For any rational number we consider the representation
$$
x=p^{\gamma}{m\over n}
$$
where $p$ is a prime, $\gamma$ is an integer, $p$, $m$, $n$ are
mutually prime numbers, $n\ne 0$. The $p$--adic norm is defined as
follows
$$
|x|_p=p^{-\gamma}
$$
A $p$--adic number can be uniquely represented by the series
$$
x=\sum_{j=\gamma}^{\infty}x_jp^j,\qquad x_j=0,\dots,p-1
$$
which converges in the $p$--adic norm.

A complex valued character $\chi:Q_p\to C$ of a $p$--adic argument
(where $x$ has the form of the above series) is defined by
$$
\chi(x)=\exp \left(2\pi i \sum_{j=\gamma}^{-1}x_j p^j\right),\qquad
\chi(x+y)=\chi(x)\chi(y)
$$
When $\gamma$ above is nonnegative the sum above is equal to zero
and the character is equal to one.

The character $\chi$ is a locally constant function. The function
$f$ is locally constant if for any $x$ there exists $\varepsilon$
for which $\forall y$: $|x-y|_p\le \varepsilon$ we have $f(x)=f(y)$.

Another example of a locally constant function is characteristic
function of a $p$--adic ball. We denote $\Omega(x)$ the
characteristic function of the interval $[0,1]$.  The characteristic
function of the $p$--adic ball of radius 1 with center in 0 has the
form:
$$
\Omega(|x|_p)=\left\{
\begin{array}{rc}
1,&|x|_p\le 1\;,\\
0,&|x|_p>1\;.\\
\end{array}
\right.
$$
Let us note that, unlike in the real case, the characteristic
function of a $p$--adic ball is continuous (the same holds for an
arbitrary locally constant function).

The Bruhat--Schwartz space ${\cal D}(Q_p)$ of $p$--adic test
functions is the linear space of locally constant complex valued
functions with compact support. Any function in ${\cal D}(Q_p)$ is a
(finite) linear combination of characteristic functions of balls.

The Vladimirov operator of $p$--adic fractional differentiation has
the following form
$$
D^{\alpha} f(x)=\frac{1}{\Gamma_p(-\alpha)}
\int_{Q_p}\frac{f(x)-f(y)}{|x-y|_p^{1+\alpha}}d\mu(y),\qquad
\alpha>0
$$
Here $\mu$ is the Haar measure on $p$--adic field (for which the
measure of a ball is equal to the diameter of this ball). The
$p$--adic gamma function has the form
$$
\Gamma_p(-\alpha)={p^{\alpha}-1\over 1-p^{-1-\alpha}}
$$
The domain of $D^{\alpha}$ contains the space ${\cal D}(Q_p)$.

The Vladimirov operator generates a natural random walk on the
$p$--adic field considered in \cite{RandomWalk}. For more general
considerations of $p$--adic dynamics see \cite{Andr}.

The basis of discrete $p$--adic wavelets and its relations to the
spectral theory of $p$--adic pseudodifferential operators is
described in the following theorem \cite{wavelets}:

\begin{theorem}\label{wav1}\qquad {\sl \noindent 1) The set of functions $\{\psi_{\gamma
nj}\}$:
\begin{equation}\label{basispaw}
\psi_{\gamma nj}(x)=p^{-{\gamma\over 2}} \chi(p^{\gamma-1}j
(x-p^{-\gamma}n)) \Omega(|p^{\gamma} x-n|_p),
$$
$$
x\in Q_p,\quad\gamma\in {\bf Z},\quad n\in Q_p/Z_p,\quad
j=1,\dots,p-1
\end{equation}
is an orthonormal basis in $L^2(Q_p)$.

\noindent 2) This basis consists of eigenvectors of the Vladimirov
operator $D^{\alpha}$:
\begin{equation}\label{eigenvalues}
D^{\alpha}\psi_{\gamma nj}= p^{\alpha(1-\gamma)}\psi_{\gamma nj}
\end{equation}
here $n\in Q_p/Z_p$ in (\ref{basispaw})  are represented by the
numbers \be\label{nQpZp} n=\sum_{l=\beta}^{-1}n_l p^l,\qquad
n_l=0,\dots,p-1 \ee }

\end{theorem}

It is possible to compute, using the basis of $p$--adic wavelets,
spectra of a very wide class of $p$--adic pseudodifferential
operators \cite{nhoper}. Multidimensional $p$--adic wavelets in
relation to spectral theory of $p$--adic pseudodifferential
operators were discussed in \cite{AKhSh}. Wavelets on some family of
abelian locally compact groups were considered in \cite{Benedetto}.
There exists a generalization of theory of $p$--adic wavelets and
pseudodifferential operators onto general locally compact
ultrametric spaces \cite{Izv}, \cite{ACHA}, \cite{MathSbornik}.

There exists a natural map of $p$--adic numbers onto the positive
half--line $R_{+}=[0,\infty)$, which, for $p=2$, maps the basis of
2--adic wavelets onto the basis of Haar wavelets on the half--line.
This exhibits a natural relation between real and $p$--adic wavelet
theories.

Let us discuss this relation. The wavelet basis in $L^2({ R})$
contains the functions
$$
\Psi_{\gamma  n}(x)=2^{-{\gamma\over 2}}\Psi(2^{-\gamma}x-n),\quad
\gamma\in { Z},\quad n \in { Z}
$$
where $\Psi$ is some integrable mean zero function. The simplest
example is the Haar wavelet
$$
\Psi(x)=\chi_{[0,\frac{1}{2}]}(x)- \chi_{[\frac{1}{2},1]}(x)
$$
equal to the difference of two characteristic functions.

Define the $p$--adic change of variable (or the Monna map) as
follows:
$$
\rho:Q_p \to { R}_+
$$
\be\label{padicchange} \rho:\sum_{i=\gamma}^{\infty} x_i p^{i}
\mapsto \sum_{i=\gamma}^{\infty} x_i p^{-i-1},\quad
x_i=0,\dots,p-1,\quad \gamma \in Z \ee

This map is one-to-one almost everywhere and conserves the measure:
$$
\mu(S)=L(\rho(S))
$$
where $S$ is a measurable set, $\mu$ is the $p$--adic Haar measure,
$L$ is the Lebesgue measure on the half--line. The map $\rho$
satisfies the H\"older inequality:
$$
|\rho(x)-\rho(y)| \le |x-y|_p
$$

The $p$--adic change of variable is a one--to--one map between the
group $Q_p/Z_p$ and the set $N_0$ of natural numbers including zero.
Here the elements of $Q_p/Z_p$ are given by the rational numbers
(\ref{nQpZp}).

The following lemma and theorem were proven in \cite{wavelets}.

\begin{lemma}\label{ballmaps}\qquad {\sl For $n\in Q_p/Z_p$ and $m,k\in Z$, $k\ge m$ the map
$\rho$ satisfies the conditions
\begin{equation}\label{in}
\rho: p^m n+p^k Z_p\to p^{-m} \rho(n)+[0,p^{-k}]
\end{equation}
\begin{equation}\label{out}
\rho:Q_p\backslash \{p^m n+p^k Z_p\}\to R_+\backslash\{ p^{-m}
\rho(n)+(0,p^{-k})\},\qquad n\ne 0
\end{equation}
$$
\rho:Q_p\backslash \{p^k Z_p\}\to R_+\backslash [0,p^{-k})
$$
}

\end{lemma}

\begin{theorem}\label{wav2}\qquad{\sl For $p=2$ the map
(\ref{padicchange}) provides a one to one correspondence between the
Haar basis in $L^2(R_+)$ and the basis of $p$--adic wavelets in
$L^2(Q_p)$: \be\label{mapswavelets} \Psi_{\gamma\rho(n)}(\rho(x))=
\psi_{\gamma n1}(x) \ee }
\end{theorem}

We understand formula (\ref{mapswavelets})  in the $L^2$ sense
(since the map $\rho$ is not a one-to-one map on the set of measure
zero).

One can consider the continuous wavelet transform on the field of
$p$--adic numbers \cite{Altaisky}. The affine group on the field of
$p$--adic numbers in the unitary representation $G$ in $L^2(Q_p)$
acts by translations and dilations
$$
G(a,b)f(x)=f^{a,b}(x)=|a|_p^{-{1\over 2}}f\left({x-b\over a}\right)
$$
where $a,b\in Q_p$  and $a\ne 0$.

Assume we have a wavelet on $Q_p$ --- i.e. a mean zero  complex
valued function $\Psi$ in $L^1(Q_p)\bigcap L^2(Q_p)$. The wavelet
transform $Tf$ of the function $f\in L^2(Q_p)$ is defined by
$$
(Tf)(a,b)=|a|_p^{-{1\over 2}} \int dx f(x)\ov{\Psi}\left({x-b\over
a}\right)=\la \Psi^{a,b},f\ra
$$
where $\la\cdot,\cdot\ra$ is the scalar product in $L^2(Q_p)$.

The inverse wavelet transform has the form \be\label{inv_wav}
f(x)=C_{\Psi}^{-1}\int{dadb\over |a|_p^2}\la \Psi^{a,b},f\ra
\Psi^{a,b}(x)=C_{\Psi}^{-1}\int{dadb\over |a|_p^2}(Tf)(a,b)
\Psi^{a,b}(x) \ee Here
$$
C_{\Psi}={1\over\|\Psi\|^2}\int{dadb\over
|a|_p^2}|\la\Psi(x),\Psi^{a,b}(x)\ra|^2
$$
and $\|\Psi\|$ is the $L^2$--norm.

The formula for the inverse wavelet transform (\ref{inv_wav}) has
the equivalent form of an expansion of the unit
$$
C_{\Psi}^{-1}\int{dadb\over |a|_p^2} |\Psi^{a,b}\ra \la
\Psi^{a,b}|=1
$$
Here we use the standard notations from quantum mechanics:
$|\Psi^{a,b}\ra$ denotes the element of the Hilbert space $L^2(Q_p)$
given by the function $\Psi^{a,b}$, $\la \Psi^{a,b}|$ is the
canonically conjugate linear bounded functional on $L^2(Q_p)$, and
$|\Psi^{a,b}\ra \la \Psi^{a,b}|$ is the orthogonal rank one
projection onto $\Psi^{a,b}\in L^2(Q_p)$.

\bigskip\bigskip

\noindent{\bf Acknowledgments}\qquad One of the authors (S.K.) would
like to thank I.V.Volovich and V.S.Vla\-di\-mi\-rov for fruitful
discussions and valuable comments. He gratefully acknowledges being
partially supported by the grant DFG Project 436 RUS 113/809/0-1, by
the grants of The Russian Foundation for Basic Research  RFFI
05-01-04002-NNIO-a and RFFI 05-01-00884-a, by the grant of the
President of Russian Federation for the support of scientific
schools NSh 6705.2006.1 and by the Program of the Department of
Mathematics of Russian Academy of Science ''Modern problems of
theoretical mathematics''.

\end{document}